\documentclass[12pt,twoside]{article}
\usepackage{fleqn,espcrc1}

\usepackage{graphicx}

\def\beq{\begin{equation}}
\def\eeq{\end{equation}}
\def\bea{\begin{eqnarray}}
\def\eea{\end{eqnarray}} 
\def\eqlab#1{\label{eq:#1}}
\def\figlab#1{\label{fig:#1}}

\def\eref#1{(\ref{eq:#1})}
\def\eqref#1{eq.~(\ref{eq:#1})}
\def\Eqref#1{Eq.~(\ref{eq:#1})}

\def\Figref#1{Fig.~\ref{fig:#1}}

\def\sla#1{#1 \!\! \slash}

\def\slap{p \hspace{-1.8mm} \slash}

\def\quarter{\mbox{\small{$\frac{1}{4}$}}}
\def\third{\mbox{\small{$\frac{1}{3}$}}}

\def\al{\alpha}

\def\ga{\gamma} \def\Ga{{\it\Gamma}}
 \def\De{\Delta}
\def\veps{\varepsilon}  

\def\la{\lambda} \def\La{{\it\Lambda}}

\def\si{\sigma} 
  
\def\w{\omega}

\def\pa{\partial}
\def\vrho{\varrho}

\def\pa{\partial}

\def\nn{\nonumber}

\def\lag{{\mathcal L}}

\def\psib{\overline\psi}

\def\ol#1{\overline{#1}}

\def\CF#1#2#3#4{#1 {\bf #2},  #4 (#3)}  
\def\ibid {{\it ibid.}}

\def\prd {Phys.\ Rev.\ D}

\newcommand{\AmS}{{\protect\the\textfont2
  A\kern-.1667em\lower.5ex\hbox{M}\kern-.125emS}}

\title{Electromagnetic moments of relativistic higher spin baryons}

\author{V. Pascalutsa\thanks{Supported by the 
Australian Research Council (ARC).}\\[3mm]
Department of Physics, Flinders University, 
Bedford Park, SA 5042, Australia %
}
       
\begin{document}

\newsavebox{\pinb}
\sbox{\pinb}{\parbox{12cm}{
\resizebox{9cm}{!}{\includegraphics*[2.5cm,21cm][15cm,23.5cm]{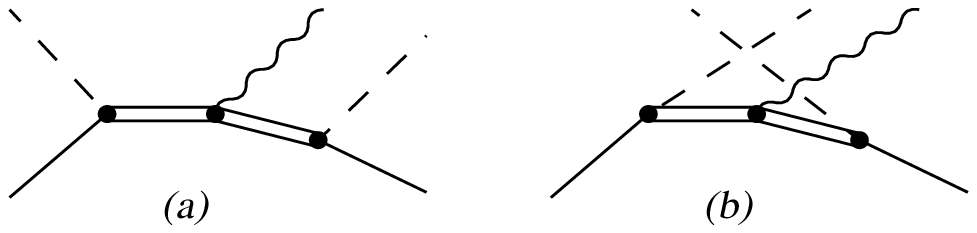} }} }

\newsavebox{\ping}
\sbox{\ping}{\parbox{12cm}{
\resizebox{7cm}{!}{\includegraphics*[2cm,18.2cm][19cm,23.5cm]{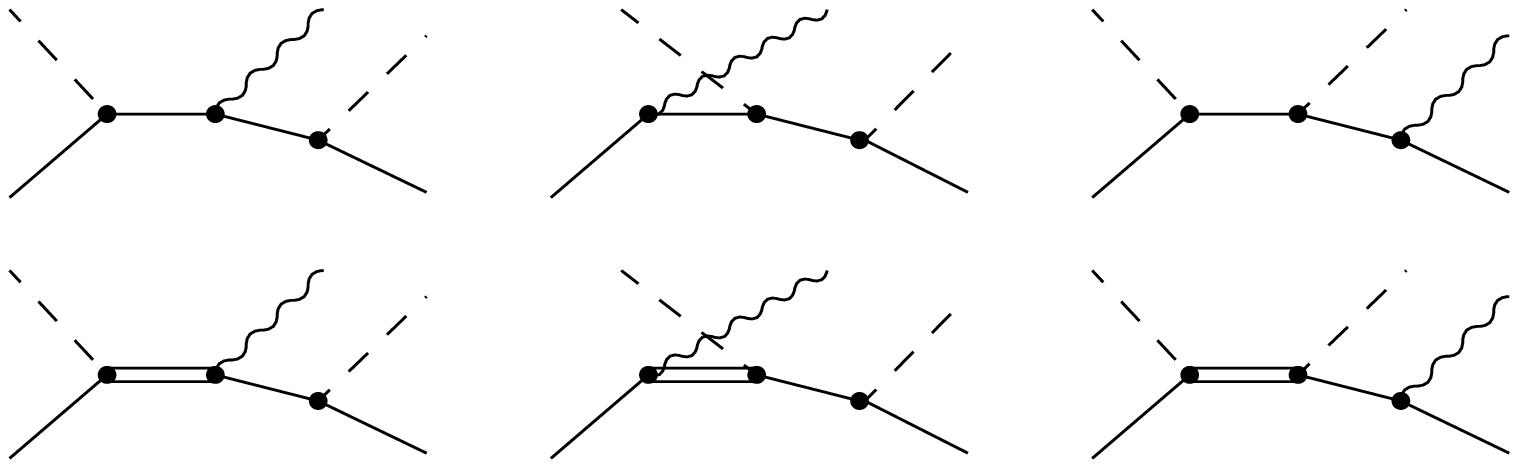} }} }

\maketitle

\begin{abstract}
We point out a source of ambiguities
in the measurements of the electromagnetic moments 
of spin-3/2 baryons which rely on relativistic Lagrangian
models. An anambiguous relation between the parameters
of the spin-3/2 electromagnetic Lagrangian and 
the electromagnetic moments of the spin-3/2 particle
exits in general only for on-shell situation, while 
the measurements are done on ``virtual'' baryons. 
\end{abstract}

\bigskip

Accurate measurement of electromagnetic (EM) moments of $N^*$ 
resonances could provide an important testing ground for
many quark-models and lattice QCD
predictions. Because of the short lifetime of the
$N^*$ resonances such measurements present experimental as
well as theoretical challenge. At present one can hope only
for indirect measurements. For example,
the dipole magnetic moment of the $\Delta^{++,0}$ isobars 
is extracted from the observables of radiative pion-nucleon 
scattering, $\pi^{\pm}+p\rightarrow\pi^{\pm}+p+\gamma$ 
(see, e.g., \cite{Bos91}), 
while the magnetic moment of the $\Delta^+$ can possibly soon 
be measured at MAMI in the radiative pion 
photoproduction \cite{Met00}, 
$\gamma+p\rightarrow\pi^0+p+\gamma$.

In these reactions the $\ga \De\De$ current containing
the EM moments is involved
only through more complicated mechanisms, see 
\Figref{graph1}. Moreover, other mechanisms
(such as in \Figref{graph2})
may interfere and must be estimated reliably.
The necessary theoretical input is usually provided
by some effective Lagrangian model based on hadronic
degrees of freedom~\cite{pinbrlg,Mac99,Dre00}, where 
all the relevant mechanisms can be computed, while
the parameters other than the EM moments
can be fixed from different sources (e.g.,
by using the same effective Lagrangian for the
description of other processes). 
 
Here we shall assume a general form of the effective 
hadronic Lagrangian restricted only by a few
fundamental principles such as Lorentz and gauge invariance,
and possibly the conditions due to the approximate chiral symmetry.

We will consider the charge and dipole magnetic moment interactions
 of the spin-3/2 baryon and show how the parameters
of the Lagrangian can unambiguously be related to the
magnetic moment of the ``real'' spin-3/2 particle.
For virtual particle this relation is in general ambiguous.
This problem is also related to the
so-called ``off-shell ambiguities'' of the covariant 
spin-3/2 description which generally does not exclude off-shell
the unphysical spin-1/2 contributions. (Recently,
Machavariani {\em et al.}~\cite{Mac99} have claimed to get rid of
the ``off-shell ambiguities'' formulation by writing 
the spin-3/2 propagator in terms of on-shell Rarita-Schwinger vector-spinors.
However, in doing so one looses relativistic covariance, see
Ref.~\cite{FaT84} for details.) 
Furthermore, by considering the matrix elements of
processes such as in \Figref{graph1}, we can argue that certain
consistency conditions on the form of the
$\pi N\De$ and $\ga N\De$  couplings
may allow us to deal with the ``off-shell ambiguities''. 

\begin{figure}  
\begin{minipage}[t]{75mm}
\usebox{\pinb}
\caption{Possible mechanisms of $\pi N$ bremsstrahlung involving the EM
moments of the $\De$.}
\figlab{graph1}
\end{minipage}
\hspace{\fill}
\begin{minipage}[t]{80mm}
\usebox{\ping}
\caption{Some other relevant mechanisms of  $\pi N$ bremsstrahlung.}
\figlab{graph2}
\end{minipage}
\vskip-5mm
\end{figure}%

To describe the spin-3/2 fields of (decuplet) baryons
it is natural to use the covariant Rarita-Schwinger (RS)
formalism\cite{Rar41} where the field is a Lorentz vector-spinor
$\psi^\mu (x)$ with the following free Lagrangian\footnote{Other
forms of free spin-3/2 Lagrangian are frequently used in the literature,
but they relate to \Eqref{rs} by a field redefinition
$\psi_\mu \rightarrow (g_{\mu\nu}+b \ga_\mu\ga_\nu) \psi^\nu$, with 
$b\neq -\quarter$. Present discussion is fixed to representation
\eref{rs}, however is unchanged for other choices albeit the
field redefinition is done in the whole Lagrangian.} :
\beq
\eqlab{rs}
\lag_{\rm free} = \ol\psi_\mu\ga^{\mu\nu\al}\pa_\al \psi_\nu - 
m\,\ol\psi_\mu\ga^{\mu\nu} \psi_\nu
\eeq
where $m$ is the mass; $\ga^{\mu\nu}$ and $\ga^{\mu\nu\al}$ is
the totally antisymmetrized product of, respectively, 
two and three gamma-matrices. This Lagrangian leads to
the well-known RS propagator:
\beq
S^{\mu\nu}(p)=\frac{\slap+m}{p^2-m^2+i\varepsilon}
\left[-g^{\mu\nu}+\third\ga^\mu\ga^\nu
+\frac{1}{3m}(\ga^\mu p^\nu -\ga^\nu p^\mu)
+ \frac{2}{3m^2} p^\mu p^\nu\right],
\eeq
and the following free-field equations:
\beq
\eqlab{onshell}
(i\pa\!\!\! / -m )\,\psi_\mu =0\,,\,\,\,
\ga^\mu\psi_\mu =0\,,\,\,\,
 \pa^\mu\psi_\mu =0.
\eeq

Consider now the EM properties of this field. 
The electric charge is included via
minimal substitution, $\pa_\al \rightarrow 
\pa_\al + ie A_\al$, into \Eqref{rs}. To explore
the dipole magnetic moment we should include all the
nonminimal, linear in $F_{\mu\nu} = \pa_{[\mu} A_{\nu]}$
terms. We then have:
\bea
\lag_{\rm int}&=& e\left\{ i\ol\psi_\mu\ga^{\mu\nu\al} \psi_\nu A_\al
+l_1\psib{}_\mu F^{\mu \nu}\psi_\nu
+l_2\psib{}_\mu \ga. F.\ga\,\psi^\mu
+l_3 F^{\mu \nu}\,[\psib{}_\mu  \ga_\nu \,\ga.\psi+\psib.\ga \,\ga_\mu 
\psi_\nu] \right. \nn\\
&& \left.
+l_4\,\psib.\ga\, (\ga. F.\ga ) \, \ga.\psi
+i l_5\, F^{\mu \nu}\,[\psib{}_\mu  \ga_\nu \,\ga.\psi-
\psib.\ga \,\ga_\mu \psi_\nu] \right\}
\equiv  \ol\psi_\mu \La^{\mu\nu} \psi_\nu.
\eqlab{nm}
\eea
It is further argued in \cite{DPW00} that $\La^{\mu\nu}$
must be antisymmetric in order to maintain the correct
number of spin degrees-of-freedom\footnote{This condition
does not remove the consistency problem of the minimal 
coupling~\cite{JSVZ}, but only insures the proper 
degrees-of-freedom count.}: $2s+1=4$.
Hence, one finds 
$l_2=-l_4$, $l_3=-2l_4$,
and the couplings reduce to
\beq
\La^{\mu\nu}= ie \, \ga^{\mu\nu\al} A_\al
+ e m^{-1} \left[ L_1
F^{\mu\nu} +
i L_2 \ga^5\tilde{F}^{\mu\nu}  
+iL_3\,( F^{\mu \al}\,\ga_\al \,\ga^\nu -
\ga^\mu \ga_\al F^{\al\nu})\right]\, ,
\eqlab{nm1}
\eeq
where $\tilde{F}^{\mu\nu}=\veps^{\mu\nu\vrho\si}\pa_\vrho A_\si$;
$L_i$ are dimensionless parameters: 
$L_1=m (l_1+2l_2)$, $L_2=-2 m l_2$, $L_3=m l_5$.
To relate these parameters to the physical magnetic moment
we match this field theory to the soft-photon theorems.

Consider first the on-shell electromagnetic
vertex [obtained from \eref{nm} with \eref{nm1}]:
\bea
\veps_\la \Ga^\la(q) &=& e m^{-1}\,\veps_\la \,
\ol u_{\mu}(p') \left\{ -m \ga^{\mu\nu\la}
+ L_1 (q^\mu g^{\la\nu} - q^\nu g^{\la\mu}) +
i L_2 \ga^5\,\veps^{\mu\nu\si\la}  q_\si \right. \nn\\   
&& +\left. i L_3\,\left[ (q^{\mu} \ga^\la - \sla{q} g^{\la\mu})\ga^\nu  +
\ga^\mu (q^{\nu} \ga^\la - \sla{q} g^{\la\nu})\right]\right\} u_{\nu}(p)\,,
\eea
where $\veps$ and $q=p'-p$ is the polarization and the 4-momentum of
the photon; $u$ are the on-shell
RS vector-spinors, solutions of \Eqref{onshell} in momentum space.

Using the properties of the on-shell vector-spinor one can see
that (i) $L_3$ term drops out, (ii) $L_2$ term contributes only to higher
order in $q$ (basically because $\ga^5$ involves the 
``small components'' of the spinors). Also one can 
(iii) use various Gordon identities and (iv) verify
the following nontrivial identity:  
\beq
2\, \ol u'_\vrho\,\ga_{\mu\nu}\,u_\si = 
 \ol u'_{\vrho}\, ( \ga_{\mu\si} u_{\nu}  +
\ga_{\si\nu} u_{\mu}) + (\ol u'_{\mu}\ga_{\vrho\nu}
+  \ol u'_{\nu}\ga_{\mu \vrho}) u_\si + {\cal O}(q),
\eqlab{shell}
\eeq
where $u'\equiv u(p')$, $u\equiv u(p)$, $q=p'-p$. 
Using all that, the vertex to the 1st order in $q$ reads
\beq
\veps_\la \Ga^\la =  \frac{e}{2m} \, \veps_\la\,\ol {u}_{\mu}' 
\left[ (p'+p)^\la g^{\mu\nu} -
g  M^{\la\si,\,\mu\nu}_{3/2} q_\si  
\right] u_{\nu},
\eeq
with $M_{3/2}$ being the spin-3/2 Lorentz generator and 
\beq
\eqlab{gyro}
g=\frac{2}{3}(1-2 L_1).
\eeq 
On the other hand, 
according to the soft-photon theorem $g= (\mu/s)(e/2m)^{-1}$ 
is the gyromagnetic ratio; $\mu$ is the magnetic moment.\footnote{Let 
as remark on the definition of the ``anomalous magnetic
moment'', $\mu_A $. It is usually defined as
the deviation from the value implied by the minimal
coupling: $\mu_A=\mu-e/2m$. However, taking the point of view that 
$g=2$ is the universal value for all elementary (pointlike) particles with
spin~\cite{Wei70}, it is more correct to define it as
the deviation from the fundamental value, i.e., as follows:
 $\mu_A=\mu-sem^{-1}$.}

The same relation is obtained by analyzing
the Compton scattering on spin-3/2
particle. For the forward Compton amplitude
we obtain\cite{DPW00} (to the 1st order in photon frequency $\w$):
\beq
T_{fi}=\frac{e^2}{m} \,\veps'.\veps \, \ol u'. u +
\frac{ie^2\,\w}{4m^2}\,(g-2)^2\,
\ol u_\mu ' M^{\la\si,\,\mu\nu}_{3/2} u_\nu\, \veps_\la'\veps_\si\,,
\eeq
again with the gyromagnetic ratio given by \Eqref{gyro}. 
The use of on-shell conditions
such as \Eqref{shell} is made in obtaining this result.

Obviously, a potential problem arises when one tries to identify the
magnetic moment of the virtual particle. One cannot use any of
the statements (i)-(iv) to conclude an unambiguous relation
between the parameters of the Lagrangian and the physical
magnetic moment. In particular, because of the coupling to the
spin-1/2 sector of the RS propagator, $L_2$ and $L_3$
terms may give contributions of the same order in photon
energy as the minimal and $L_1$ terms. This problem does not arise if the
spin-3/2 propagator could be replaced by a positive-energy
spin-3/2 projection operator. However, such replacement
can be done only at the expense of loosing the Lorentz and
gauge invariance. (Gauge invariance is lost because the
EM vertex satisfies the Ward-Takahashi identity for
the full RS propagator and not for any truncated form).  

A way to deal with the problem is to 
consider the complete matrix elements (for graphs like in
\Figref{graph1}) and use the specific ``consistent''
$\pi N\De$ and $\ga N \De$ interactions~\cite{Pas98,PaT99}.
The latter interactions have the property of 
decoupling the spin-1/2 sector, and therefore
the major source of the off-shell ambiguity should be
eliminated. A study in this direction and development
of models for the radiative $\pi N$ scattering and
pion photoproduction is in progress~\cite{TiP00}.

\end{document}